\documentclass[%
preprint,
nofootinbib,
 amsmath,amssymb,
 aps,
]{revtex4-2}

\usepackage{tabularray}

\usepackage{graphicx}
\usepackage{dcolumn}
\usepackage{bm}

\begin{document}

\def\bb    #1{\hbox{\boldmath${#1}$}}

\title{ Collective behaviour in proton number fluctuations \\ seen in  $\sqrt{s_{NN}}=$ 2.4 GeV Au+Au collisions }

\author{Ali Bazgir}
\affiliation{Institute of Physics, Jan Kochanowski University, 25-406 Kielce, Poland}


\author{Maciej Rybczy\'nski}
\email{maciej.rybczynski@ujk.edu.pl}
\affiliation{Institute of Physics, Jan Kochanowski University, 25-406 Kielce, Poland}

\author{Uzair A. Shah}
\affiliation{Institute of Physics, Jan Kochanowski University, 25-406 Kielce, Poland}

\author{Zbigniew W\l odarczyk}
\affiliation{Institute of Physics, Jan Kochanowski University, 25-406 Kielce, Poland}

\begin{abstract}
At energies of a few GeV per nucleon, nuclear collisions exhibit phenomena more complex than the individualistic nucleon interactions observed at much higher energies. From recent results on proton number fluctuations in  Au + Au collisions at $\sqrt{s_{NN}}=2.4$~GeV obtained by the HADES experiment at GSI, we suggest that measuring the multiplicity distributions heavy-ion collisions can be used to probe density fluctuations associated with correlation phenomena. By using the combinant analysis, one can obtain new information contained in them and otherwise unavailable, which may broaden our knowledge of the particle interactions mechanism.
\end{abstract}
%


\maketitle
\section{Introduction}
\label{Introduction}

Recent data from the HADES Collaboration on proton number fluctuations~\cite{HADES:2020wpc} have been reported for Au+Au collisions at a nucleon pair center-of-mass collision energy of $\sqrt{s_{NN}}= 2.4$~GeV. The observed HADES data reveal significant non-Gaussian fluctuations in the number of protons within the rapidity interval $\Delta y = \pm 0.5$ around midrapidity. These pronounced fluctuations might be attributed to anomalies in the equation of state of the matter produced in the collision, potentially manifesting as local inter-proton correlations in coordinate space, possibly due to the presence of a critical point in the baryon-rich regime\cite{Vovchenko:2022szk}. However, substantial fluctuations can also arise from external global factors that apply even to a system of non-interacting particles~\cite{Savchuk:2022ljy}. Fluctuations in baryon number, especially their critical or pseudo-critical behavior, are typically characterized by moments, factorial moments, cumulants, or higher-order cumulants of the observed particle number distribution~\cite{Broniowski:2017tjq, Bzdak:2012ab}. These variables strongly depend on the experimental acceptance procedure~\cite{Savchuk:2022ljy}. Remarkably, the acceptance procedure does not change the type of multiplicity distributions (it changes only the parameters of the distributions)~\cite{Rybczynski:2019pxr}.

In the present paper, we consider an alternative possibility to obtain information on proton number distributions. The paper is organized as follows. In Sec.~\ref{formulas} we present the formulas of the binomial acceptance procedure and discuss recurrence relations for multiplicity distribution. In Sec.~\ref{results} the HADES results are analyzed and fitted within the the compound distribution. Conclusions presented in Sec.~\ref{Concl} summarize the article. 


\section{Binomial acceptance, recurrence relation, and combinants}
\label{formulas}

Let's consider that $g\left(M\right)$ represents the real distribution characterizing the multiplicity distribution across the entire phase space. However, in an experimental setting, the multiplicity can only be measured within a specific rapidity window, $\Delta y$, and a range of transverse momentum, $\Delta p_{T}$.

We can model the detection process as a Bernoulli process, which follows a binomial distribution (BD) with the generating function
\begin{equation}
F\left(z\right) = 1 - \alpha + \alpha z,
\label{F-1}
\end{equation}
where $\alpha$ is the probability that a particle will be detected within the rapidity window. The number of detected particles is given by
\begin{equation}
N = \sum_{i=1}^{M} n_{i},
\label{N-1}
\end{equation}
where each $n_{i}$ follows the BD characterized by the generating function $F(z)$, and $M$ is distributed according to $g\left(M\right)$, which has the generating function $G(z)$. Therefore, the generating function for the observed multiplicity distribution $P\left(N\right)$ is given by
\begin{equation}
H\left(z\right) = G\left[F\left(z\right)\right],
\label{H-1}
\end{equation}
leading to the multiplicity distribution
\begin{equation}
P\left(N\right) = \frac{1}{N!} \frac{d^N H\left(z\right)}{dz^{N}}\Bigg|_{z=0}.
\label{PN-1}
\end{equation}

This approach, when applied to specific distributions like the negative binomial distribution (NBD), Poisson distribution (PD), or BD, yields the same forms of distributions but with adjusted parameters. Specifically, their generating functions are
\begin{equation}
G\left(z\right) = \left\{\begin{array}{rcl}
\left[\left(1-p\right)/\left(1-pz\right)\right]^k & \mbox{for NBD} \\ 
exp\left[\lambda\left(z-1\right)\right] & \mbox{for PD} \\  
\left[1+p\left(z-1\right)\right]^K & \mbox{for BD}  
\end{array}\right.
\label{G-1}
\end{equation} 
with the parameters being modified as follows: for BD, $p \rightarrow p' = \alpha p$; for PD, $\lambda \rightarrow \lambda' = \alpha \lambda$; and for NBD, $p \rightarrow p' = \alpha p / \left[1 - p\left(1-\alpha\right)\right]$.

The underlying dynamics of multiparticle production can be captured by the relationships between successive measured multiplicities $N$. In the simplest case, it is assumed that the multiplicity $N$ is influenced only by its neighboring values $\left(N \pm 1\right)$ through a straightforward recurrence relation:
\begin{equation}
\left(N+1\right) P\left(N+1\right) = r\left(N\right) P\left(N\right).
\label{rec-1}
\end{equation}
A commonly used form of $r\left(N\right)$ is linear, where $r\left(N\right) = a + bN$. In this context, $b=0$ corresponds to PD, $b>0$ to NBD, and $b<0$ to BD.

Experimental observations often indicate that the measured $P\left(N\right)$ contains additional information not fully accounted for by the recurrence relation in Eq.~(\ref{rec-1}), which may be overly restrictive. In~\cite{Rybczynski:2018bwk}, we proposed a more general recurrence relation, which is used in counting statistics for point processes with multiplication effects. Unlike Eq.~(\ref{rec-1}), this generalized form connects all multiplicities through a set of coefficients $C_{j}$, defining the corresponding $P\left(N\right)$ as follows:
\begin{equation}
\left(N+1\right)P\left(N+1\right)=\langle N\rangle \sum^{N}_{j=0} C_{j} P\left(N - j\right). \label{rr}
\end{equation}
These coefficients $C_{j}$, known as combinants, carry the memory of the $(N+1)^{th}$ particle with respect to all previously produced $N-j$ particles. They can be directly calculated from the experimentally measured $P\left(N\right)$ by inverting Eq.~(\ref{rr}) into the following recurrence formula for combinants $C_{j}$~\cite{Rybczynski:2018bwk}:
\begin{equation}
\langle N\rangle C_{j} = \left(j+1\right)\left[\frac{P\left(j+1\right)}{P\left(0\right)}\right] - \langle N\rangle \sum^{j-1}_{i=0}C_{i} \left[\frac{P\left(j-i\right)}{P\left(0\right)}\right]. \label{rCj}
\end{equation}
Combinants $C_{j}$ can also be expressed using the generating function $G\left(z\right)$ of $P\left(N\right)$ as follows:
\begin{equation}
\langle N\rangle C_{j}= \frac{1}{j!} \frac{d^{j+1} \ln G\left(z\right)}{dz^{j+1}}\Bigg|_{z=0}.
\label{Cj}
\end{equation}


\section{HADES results for proton number fluctuations}
\label{results}

In this paper, we analyze the HADES experiment data on proton-number fluctuations measured in Au+Au collisions at a center-of-mass energy of $\sqrt{s_{NN}}=2.4$~GeV~\cite{HADES:2020wpc}. The HADES experiment is a fixed-target setup located at the heavy-ion synchrotron SIS18 at the GSI Helmholtz Centre for Heavy Ion Research in Darmstadt, Germany. The Au+Au reactions were conducted using a stack of 15 gold pellets, each 25 $\mu$m thick, in total 0.375~mm, which corresponds to a nuclear interaction probability of 1.35\%. A gold beam with a kinetic energy of $E_{kin}=1.23$~GeV per nucleon and an average intensity of $1-2\cdot 10^{6}$ particles per second was directed at the gold target. To ensure uniform and symmetric acceptance around midrapidity, the phase-space region for the HADES data on proton-number fluctuations spans a laboratory rapidity of $y=y_{0}\pm 0.5$ and transverse momentum range of $0.4 < p_{T} < 1.6$~GeV$/c$, where midrapidity $y_{0} = 0.74$~\cite{HADES:2020wpc}.

\begin{figure}
\begin{center}
\includegraphics[scale=0.95]{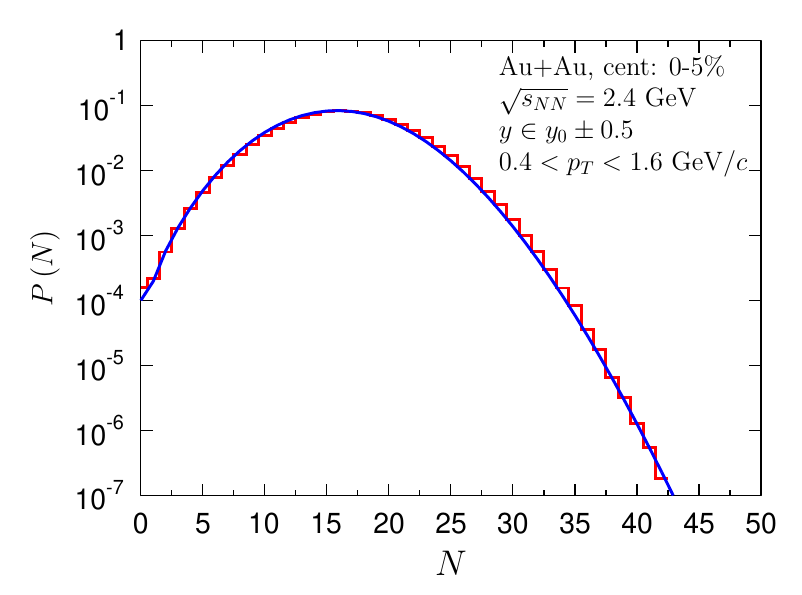}
\end{center}
\vspace{-10mm}
\caption{Proton multiplicity distribution extracted from HADES experimental data~\cite{HADES:2020wpc} (histogram) and fitted by the compound CBD distribution~\ref{comp-3} (line).}
\label{FIG-1}
\end{figure}

\begin{figure}
\begin{center}
\includegraphics[scale=0.95]{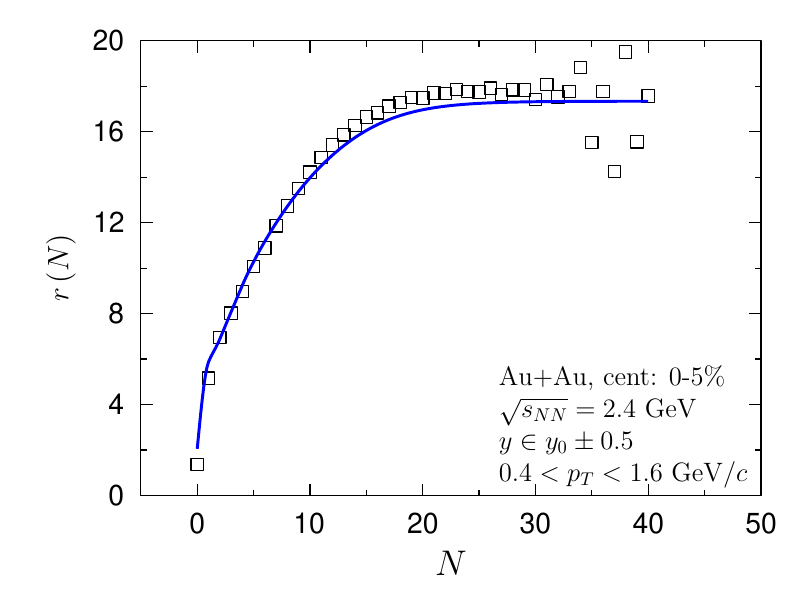}
\end{center}
\vspace{-10mm}
\caption{Recurrence relation $r\left(N\right)=\left(N+1\right)P\left(N+1\right)/P\left(N\right)$ evaluated from HADES experimental data~\cite{HADES:2020wpc} presented on Fig.~\ref{FIG-1} (squares) and our compound model (line).}
\label{FIG-2}
\end{figure}

\begin{figure}
\begin{center}
\includegraphics[scale=0.95]{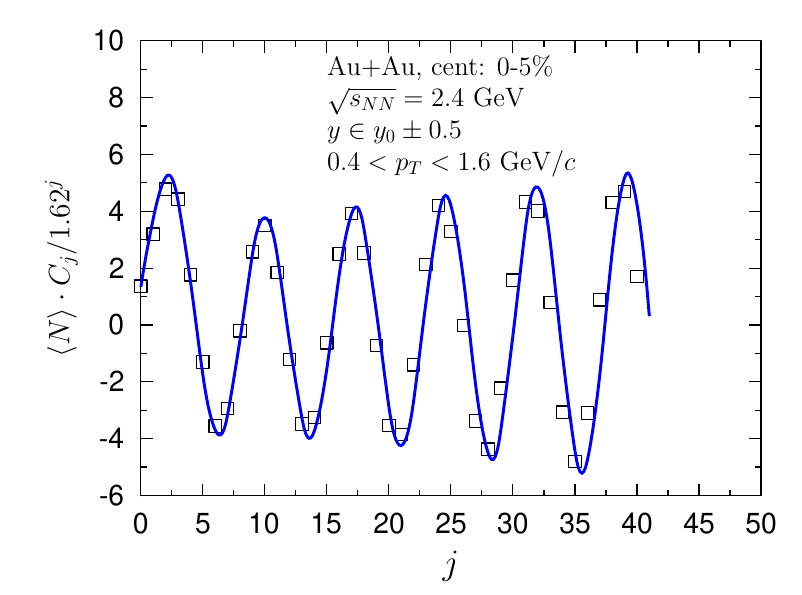}
\end{center}
\vspace{-10mm}
\caption{Combinants $\langle N\rangle C_{j}$ emerging from $P\left(N\right)$ distribution presented on Fig.~\ref{FIG-1} fitted by our compound model.}
\label{FIG-3}
\end{figure}

\begin{table}
\centering
\caption{Mean values $\langle N\rangle$ and scaled variances $\omega=Var\left(N\right)/\langle N\rangle$ of proton number distributions, measured in bin rapidities $\Delta y$, for Au+Au collisions at different collision  centralities. Half period $\lambda$ of combinant oscillations determine number of bunches $\langle M\rangle=\langle N\rangle/\lambda $ \label{TAB}.}
\begin{tblr}{
  cell{2}{1} = {r=8}{},
  cell{10}{1} = {r=8}{},
  vlines,
  hline{1-2,10,18} = {-}{},
  hline{3-9,11-17} = {2-6}{},
}
$\Delta y$  & Centrality [\%] & $\langle N\rangle$     & $\omega$   & $\lambda$ & $\langle M\rangle$ \\
0.5 & 0 - 5           & 16.44 & 1.40 & 3.6 & 4.6  \\
    & 5 - 10          & 14.35 & 1.38 & 2.9 & 4.9  \\
    & 10 -15          & 12.28 & 1.38 & 2.5 & 4.9  \\
    & 15 20           & 10.39 & 1.43 & 1.9 & 5.5  \\
    & 20 -25          & 8.70  & 1.54 & 1.8 & 4.8  \\
    & 25 -30          & 7.46  & 1.59 & 1.6 & 4.7  \\
    & 30 - 35         & 6.63  & 1.54 & 1.2 & 5.5  \\
    & 35 -40          & 5.93  & 1.47 & 1.3 & 4.6  \\
0.2 & 0 - 5           & 7.60  & 1.20 & 1.7 & 4.5  \\
    & 5 - 10          & 6.52  & 1.19 & 1.2 & 5.4  \\
    & 10 - 15         & 5.47  & 1.18 & 1.2 & 4.6  \\
    & 15 - 20         & 4.55  & 1.21 & 1.1 & 4.2  \\
    & 20 - 25         & 3.75  & 1.27 & 1.0 & 3.8  \\
    & 25 - 30         & 3.16  & 1.30 & 1.0 & --   \\
    & 30 -35          & 2.78  & 1.28 & 1.0 & --   \\
    & 35 - 40         & 2.45  & 1.25 & 1.0 & --   
\end{tblr}
\end{table}

In Fig.~\ref{FIG-1}, we present a typical proton number distribution registered in HADES experiment~\cite{HADES:2020wpc}. The recurrence relation depicted in Fig.~\ref{FIG-2} reveals a highly non-linear behavior, underscoring the complex structure of $P\left(N\right)$. Since a single distribution type, such as NBD, PD, or BD, fails to adequately describe the data, we explore the concept of compound distributions. These distributions are relevant when the production process involves the generation of a certain number $M$ of objects (such as clusters, fireballs, etc.) according to a distribution $f\left(M\right)$ (with generating function $F\left(z\right)$). These objects then decay independently into multiple secondary particles $n_{i=1,2,...,M}$, each following the same distribution $g\left(n\right)$ (with generating function $G(z)$). The resulting multiplicity distribution,
\begin{equation}
h\left(N=\sum^M_{i=0}n_{i} \right)=f\left(M\right) \otimes g\left(n\right),
\label{comp-1}
\end{equation}
is a compound distribution of $f$ and $g$, with a generating function given by
\begin{equation}
H\left(z\right) = F\left[G\left(z\right)\right].
\label{comp-2}
\end{equation}

The oscillatory pattern of the combinants $C_{j}$ shown in Fig.~\ref{FIG-3} restricts the set of distributions $P\left(N\right)$ that can lead to such oscillations to, primarily, BD and any compound distributions based on BD~\cite{Rybczynski:2018bwk}. To control the oscillation period, one can combine this BD with another distribution. For instance, we use a Poisson distribution (where $C_{0}=1$ and $C_{j>0}=0$) for this purpose. The generating function of the resulting compound distribution, CBD=BD\&PD, is
\begin{equation}
H\left(z\right) = \{p\exp\left[\lambda \left(z-1\right)\right] + 1 - p\}^{K}.
\label{comp-3}
\end{equation}

In general, the oscillation period in this case is $2\lambda$, while for BD alone, the period is 2~\cite{Rybczynski:2018bwk}~\footnote{For BD with the generating function~(\ref{G-1}), the combinants~(\ref{Cj}) are given by $\langle N\rangle C_{j}=\left(-1\right)^j K\left(p/\left(1-p\right)\right)^{j+1}$. For CBD defined by Eq.~(\ref{comp-3}), the combinants are $\langle N\rangle C_{j}=\left(-1\right)^j K\lambda^{j+1}t\left(t+1\right)^{-j-1}A_{j}\left(-t\right)$ where $t=\exp\left(\lambda\right)\left(1-p\right)/p$ and $A_{j}\left(t\right)=\left(1-t\right)^{j+1}\sum _{l=0}^\infty \left(l+1\right)^{j} t^{l}$ are the Eulerian polynomials~\cite{Hirzebruch}.}. We compare this straightforward scenario (with CBD parameters: $K=4$, $p=0.911$, and $\lambda=4.335$) to experimental data in Figs.~\ref{FIG-1}-\ref{FIG-3}.

\begin{figure}
\begin{center}
\includegraphics[scale=0.95]{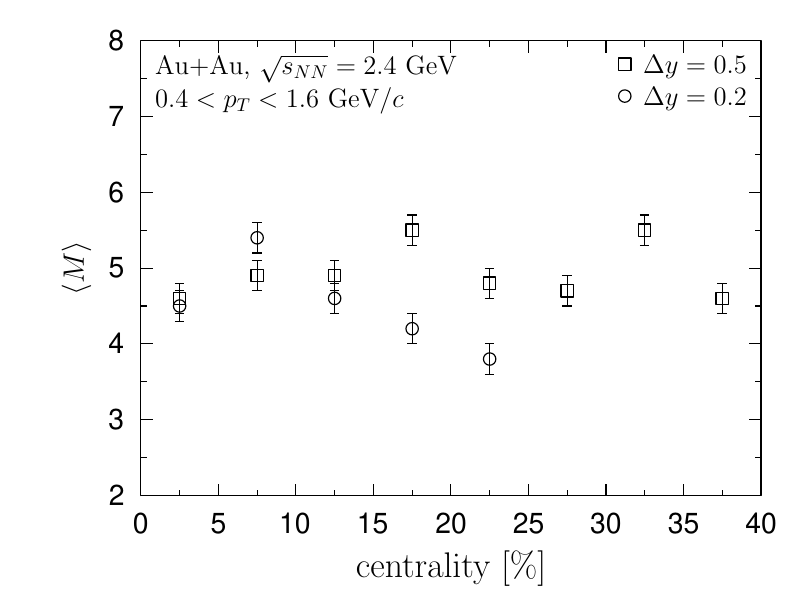}
\end{center}
\vspace{-10mm}
\caption{Number of proton bunches $\langle M\rangle=\langle N\rangle/\lambda$ evaluated from oscillations of combinants.}
\label{FIG-4}
\end{figure}

Experimental acceptance impacts the scaled variance $\omega = \text{Var}(N)/\langle N\rangle$. It's important to note that the two-particle correlation function $\langle\nu_{2}\rangle = (\omega-1)/\langle N\rangle$ remains unaffected by detector acceptance~\cite{Rybczynski:2004zi}. The mean value $\langle\nu_{2}\rangle=0.025$ is consistent for the most central collisions (in both $\Delta y=0.5$ and $\Delta y=0.2$ intervals) and increases with centrality. This trend suggests that the correlation function depends on the volume $V$ of overlapping nuclei, following $\sim\xi^{3}/V$, where $\xi$ is the correlation length. As centrality increases, the volume $V\sim\langle N\rangle$ decreases, leading to an increase in $\langle\nu_{2}\rangle$. Consequently, the scaled variance $\omega$ remains nearly constant across all centralities.

Experimental acceptance also affects the size $\lambda$, while the mean value of proton bunches number $\langle M\rangle=Kp$ is independent of acceptance. For centrality 0-5$\%$, for $\Delta y = \pm 0.5$, the size $\lambda \simeq 3.6$ and for $\Delta y = \pm 0.2$  we have $\lambda \simeq 1.7$. With increasing centrality, $\lambda$ decreases. For 0-10$\%$, $\lambda=3.5$ and for 25-30$\%$, $\lambda=1.6$ (for $\Delta y = \pm 0.5$). It is remarkable that for all $\Delta y$ bins, the number of proton bunches is roughly fixed, $\langle M\rangle=\langle N\rangle/\lambda \simeq 4.7$ with standard deviation $\sigma=0.4$.

Numerical values that characterize proton number distributions were presented in Table~\ref{TAB}. Note that for small $\langle N\rangle$ the period of oscillations is 2, i.e. the combinants are completely determined by  BD. Number of proton bunches $\langle M\rangle=\langle N\rangle/\lambda$ for different collision centralities and rapidity windows are shown in Fig.~\ref{FIG-4}.


\section{Conclusions}
\label{Concl}

The set of combinants, $C_{j}$, offers a way to measure fluctuations similar to cumulant factorial moments, $K_{q}$, which are highly sensitive to the specifics of the multiplicity distribution and have been widely used in phenomenological data analysis~\cite{Rybczynski:2018bwk}. However, combinants $C_{j}$ can provide additional insights and help address some of the unresolved questions in the field. Our approach utilizes a compound distribution (CBD), where the binomial distribution plays a central role. This binomial component is key to the oscillatory behavior observed in the combinants, while the Poisson distribution determines the period of these oscillations. In the context of cluster models, such compound distributions describe clusters using a BD and the particles within clusters with a PD. The mean multiplicity of the PD, in turn, sets the oscillation period.

In the conventional picture of a nuclear collision, the system is treated as an aggregate of essentially free nucleons. In this scheme, all observed properties are described in terms of a superposition of individual nucleon collisions. On the other hand, the virtual meson fields associated with the nucleon distribution give rise to important dynamical effects, making a Glauber treatment unfeasible. These fields tend to behave as being classical, which makes possible an approximate treatment of the field energy density using classical matter dynamics. We consider the fire-streak classical matter dynamical model. The nuclei are divided into thin tubes parallel to the beam direction (z-axis). The nuclear collision is described as a superposition of impacts between individual projectile and target tubes, associated with the same position $\left(x,y\right)$. The ratio $\langle N\rangle/\lambda$ quantifies the number $\langle M\rangle$ of such proton bunches. The transverse radius of the bunch is $R_{Au}/\sqrt{\langle M\rangle}$, equal roughly to 3 times the radius of nucleons.

Whether such a simple scenario is related to the broad spectrum of experimental data further calculations from dynamical modeling of heavy-ion collisions are required.

\vspace*{0.3cm}
\centerline{\bf Acknowledgments}
\vspace*{0.3cm}
This research was supported by the Polish National Science Centre (NCN) Grant 2020/39/O/ST2/00277. In preparation of this work, we used the resources of the Center for Computation and Computational Modeling of the Faculty of Exact and Natural Sciences of the Jan Kochanowski University of Kielce.




\end{document}